\documentclass[10pt,conference,letterpaper]{IEEEtran}

\usepackage{cite}
\usepackage{amsmath,amssymb,amsfonts}
\usepackage{graphicx}
\usepackage{textcomp}
\usepackage{booktabs}
\usepackage{multirow}
\usepackage{pifont}
\usepackage{url}

\def\BibTeX{{\rm B\kern-.05em{\sc i\kern-.025em b}\kern-.08em
    T\kern-.1667em\lower.7ex\hbox{E}\kern-.125emX}}

\begin{document}

\title{ProtoTab: Protocol-Native Tabular Representation Learning for Encrypted Traffic Classification}

\author{\IEEEauthorblockN{Sizhe Huang, Zitong Li, Shujie Yang}}

\maketitle

\begin{abstract}
Encrypted-traffic classification identifies applications and services from network flows without decrypting their contents. As encryption conceals more content and explicitly identity-bearing endpoint fields are withheld, a passive observer still sees protocol-header fields, packet order, and capture timestamps. These observations remain discriminative, but byte serialization obscures their protocol-defined field identities and temporal organization. We therefore formulate the task as protocol-native tabular representation learning: each flow is an ordered, typed packet-field table, with packets as rows, fields as columns, and flow-internal intervals along a temporal axis. ProtoTab encodes this structure through an admissible input-and-target schema, field-specific mappings, and dual-axis field/time attention. With matched observer inputs and the same raw unlabeled corpus, ProtoTab outperforms all four byte-form models across four benchmarks. It achieves 89.3\% and 83.8\% Macro-F1 on CipherSpectrum and TLS-120, exceeding the strongest matched byte baselines by 26.74 and 19.03 points. Most of each margin remains without pretraining.
\end{abstract}


\section{Introduction}\label{sec:intro}

\begin{figure}[h]
    \centering
    \includegraphics[width=\linewidth]{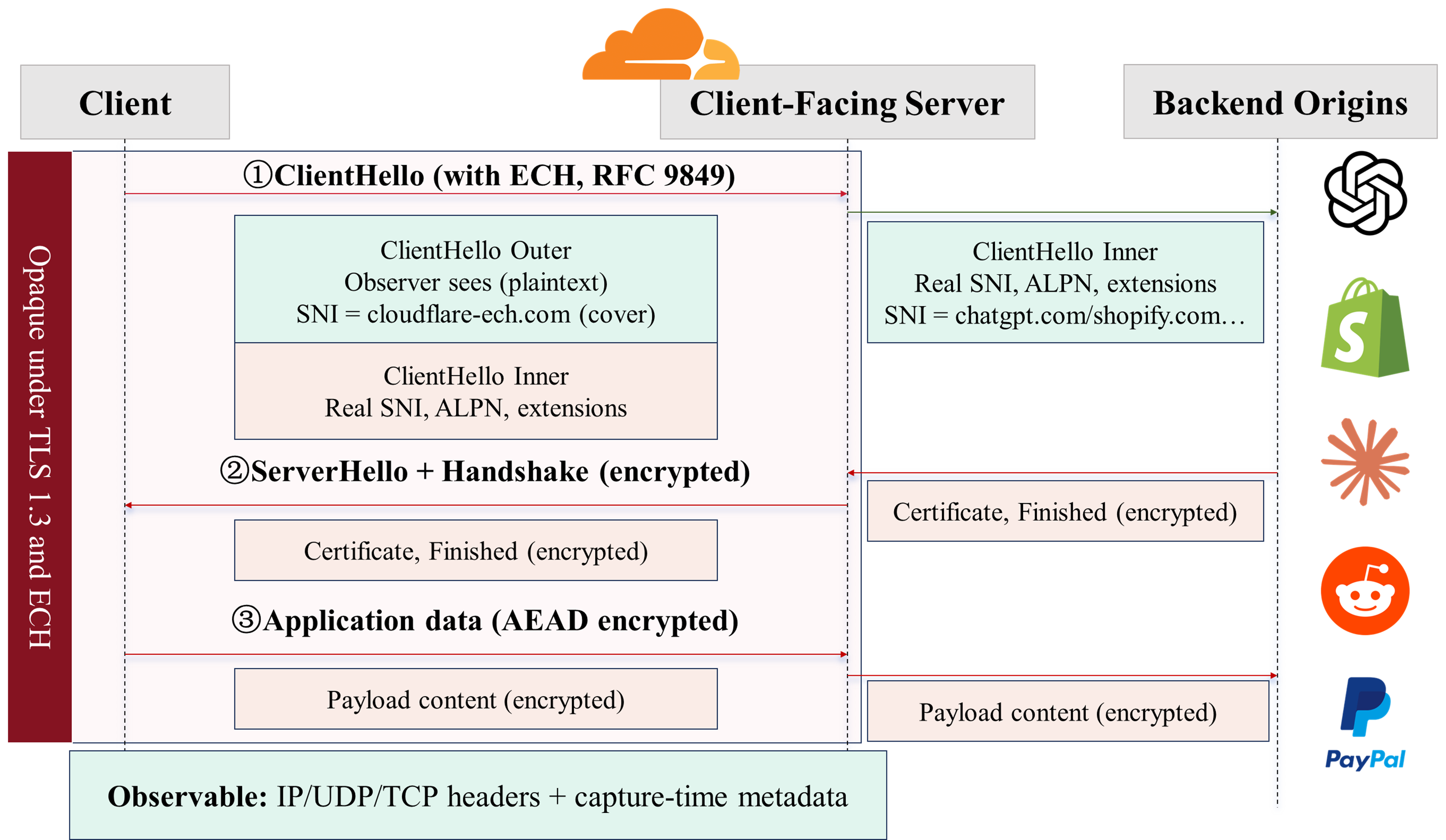}
    \caption{TLS~1.3 with ECH (RFC~9849), illustrating which signals leave passive view and which packet-level observations remain.}
    \label{fig:ech}
\end{figure}

Modern encryption reduces the information available to passive traffic classifiers. TLS~1.3 encrypts application payload content and post-ServerHello handshake messages~\cite{rfc8446,sok_etc}; ECH encrypts the ClientHelloInner carrying identifiers such as SNI and ALPN~\cite{rfc9849}; and encrypted DNS conceals name-resolution queries. Leakage-aware evaluation imposes a separate input restriction: plaintext content and identity-bearing endpoint fields should not define the task~\cite{sok_etc,pcapencoder}. After these constraints are applied, a passive observer retains generic protocol headers, packet order and direction, and capture-time metadata. Protocol parsers expose the retained headers as typed fields, while capture order and timestamps organize them across packets (Fig.~\ref{fig:ech}). This motivates our modeling question: how should these admissible observations be represented without reintroducing excluded signals?
 
\subsection{What Remains Unresolved: A Representation Gap}

Recent reassessments show that the high accuracies reported for encrypted-traffic classifiers can reflect identity-bearing or session-specific fields and packet-level train--test leakage. On their principal modern-encrypted evaluations, controlling these effects reduces the accuracy of representative models by more than 20 percentage points~\cite{sok_etc,pcapencoder}. We therefore fix the passive-observer schema and construct folds at the flow level.

\begin{figure}[h]
    \centering
    \includegraphics[width=\linewidth]{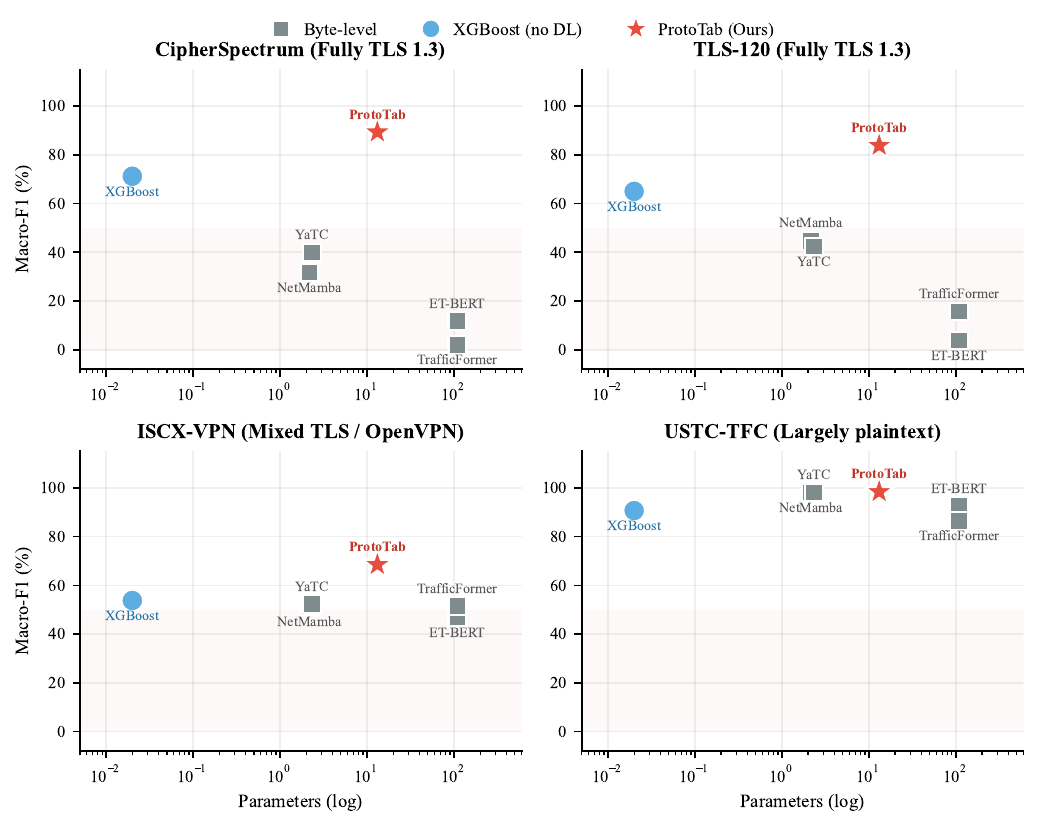}
    \caption{Retained-schema Macro-F1 versus model size; byte baselines use released checkpoints and common fine-tuning inputs.}
    \label{fig:false_prosperity}
\end{figure}

This collapse, however, does not show that the retained observer record lacks class signal. When the same retained record is exposed as a parser-defined table, XGBoost reaches 71.2\% Macro-F1 on CipherSpectrum, exceeding every evaluated byte baseline, including the strongest same-corpus result of 62.56\% (Fig.~\ref{fig:false_prosperity} and Table~\ref{tab:decomposition}). On this benchmark, the contrast rules out insufficient retained signal as the sole explanation and motivates examining how that signal is represented and learned.

\subsection{Why It Arises: Schema and Structure Mismatches}

\begin{figure}[h]
    \centering
    \includegraphics[width=\linewidth]{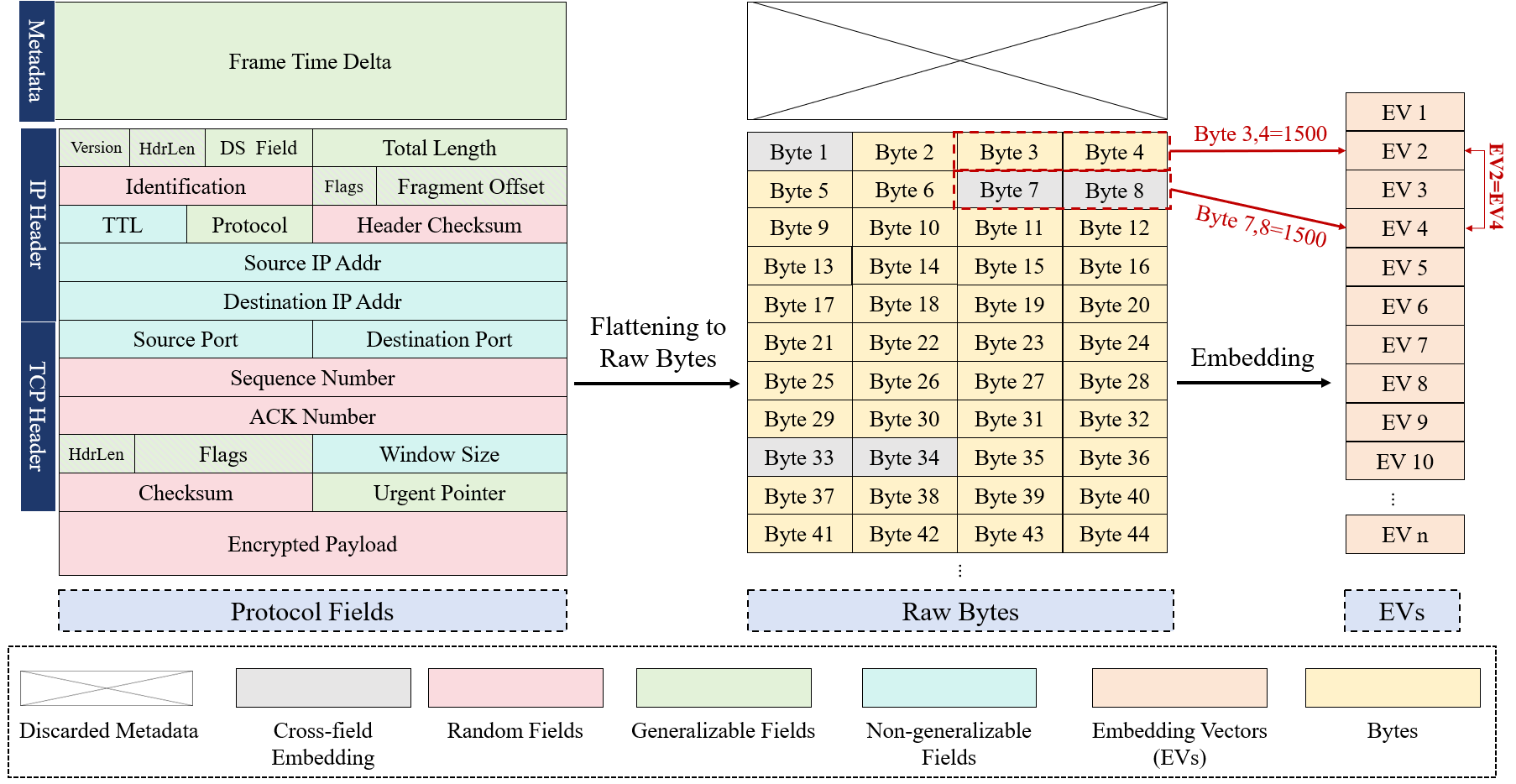}
    \caption{Three design problems addressed by ProtoTab: schema/target admissibility (P1), implicit field semantics (P2), and flattened temporal organization (P3).}
    \label{fig:violations}
\end{figure}

Byte-form models reserialize the retained observations into byte-derived layouts, leaving their parser-defined field and time organization for the learner to infer. We examine this representation sensitivity through one target-schema issue and two representation-structure issues (Fig.~\ref{fig:violations}). First, under our transfer-oriented observer model, a reconstruction target is retained only when it is observer-admissible and represents canonicalized protocol behavior; random or session-seeded values can otherwise encourage reliance on non-generalizable state (P1). Second, a shared byte vocabulary leaves protocol-field identity and value semantics implicit. Raw position is only a layout-dependent cue because variable headers and options can move a field, while identical numeric values can carry different meanings across fields (P2). Third, capture-time metadata lies outside packet bytes. In the evaluated byte-form pipelines, flow-internal inter-packet intervals are either omitted or appended as ordinary tokens, leaving their role along the packet-time axis implicit; these intervals are available online from consecutive packet arrivals (P3). The same observer record is supplied across encodings; protocol parsing and packet arrival make its field/time organization explicit.
 
\subsection{How We Address It: Protocol-Native Representation}

These principles lead to ProtoTab, which represents each flow as a typed packet-by-field table and preserves capture order as a separate temporal axis. The fixed admissibility filter enforces the same input-and-target schema (P1); fixed parser-defined columns expose field identity, one projector per column provides field-specific value mappings, and field-axis attention models within-packet relations (P2); flow-internal inter-packet intervals provide capture-time metadata, while time-axis attention models packet evolution (P3).

\textbf{Design principle.} ProtoTab changes how the retained record is organized, not what the observer can access. The observer-defined schema is fixed before learning and applies uniformly to model inputs and learning targets, while typed field identity and flow-internal timing remain explicit throughout learning and evaluation. The schema is not selected per downstream task; protocol or deployment changes therefore require a renewed admissibility audit.

\begin{figure*}[h]
    \centering
    \includegraphics[width=0.8\linewidth]{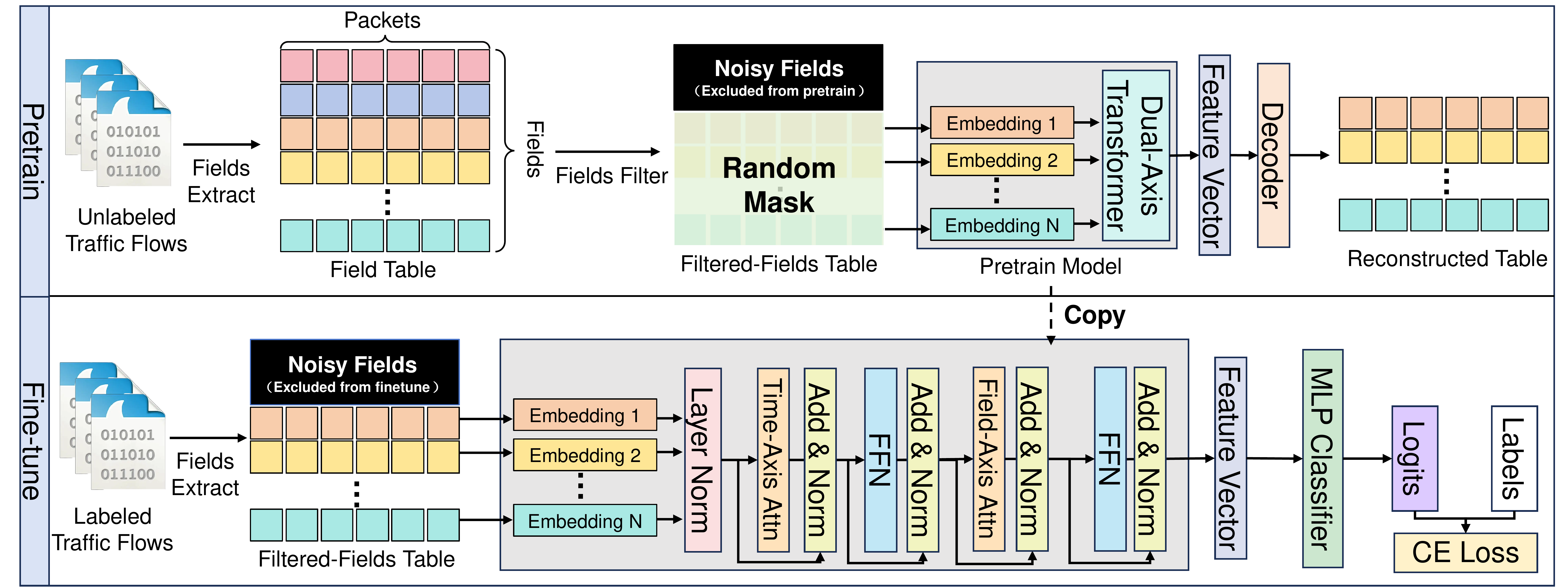}
    \caption{ProtoTab enforces the retained schema and represents traffic along field and time axes.}
    \label{fig:workflow}
\end{figure*}

\subsection{Contributions}

In summary, this paper makes the following contributions:

\begin{itemize}
    \item \textbf{Observer-Grounded Diagnosis.} We show that the retained record remains discriminative but representation-sensitive, and formulate an admissible input-and-target schema, explicit field semantics, and packet-time organization as three design principles evaluated in ProtoTab.

    \item \textbf{Protocol-Native Representation.} We introduce ProtoTab, which preserves flows as typed packet-by-field tables and separately models within-packet field relations and cross-packet evolution.

    \item \textbf{Controlled Evaluation.} We hold observer information and the raw unlabeled pretraining corpus fixed, then use matched baselines, same-schema controls, and targeted probes to measure ProtoTab's gains and examine which design choices are associated with them.
\end{itemize}

\section{Observation Model}\label{sec:observation_model}

We consider a passive network observer at a gateway, ISP tap, or campus border router. The observer captures packets in transit but cannot modify traffic and has no endpoint state, application logs, or decryption keys. The admissible model input consists of protocol headers, capture-time metadata, packet order, and packet direction. IP addresses, ephemeral (dynamic/private) ports, SNI, and ALPN are withheld; stable service ports remain as protocol-semantic fields. Payload bytes and environment-dependent fields such as TTL, TCP window size, and OS-specific options are likewise excluded following leakage-aware evaluation practices~\cite{sok_etc}.

The observer model defines retained information independently of its encoding. Consequently, byte-derived and protocol-native models are evaluated on the same retained record, encoded in their respective input forms.
\section{Related Work}

\textbf{Statistical and expert-based traffic analysis.}
Traditional ETC methods rely on expert-defined statistical summaries such as packet size distributions and flow duration~\cite{finsterbusch2013survey,uninet}. Deep Packet Inspection analyzes protocol headers and payloads but becomes ineffective under encryption~\cite{dpi}. These methods summarize flows through selected aggregate descriptors. ProtoTab instead keeps packet-level observations in table form: the same admissible parser-exposed fields are arranged as packet rows and typed protocol-field columns across tasks.

\textbf{Byte-, token-, and image-based representation learning.}
Recent representation learners adapt language, vision, and sequence models to traffic bytes or byte-derived layouts. ET-BERT~\cite{et-bert}, PERT~\cite{pert}, TrafficFormer~\cite{trafficformer}, and MIETT~\cite{miett} study token pretraining; YaTC~\cite{yatc} and other masked-vision approaches~\cite{mvm} convert packets into traffic matrices; NetMamba~\cite{netmamba} applies state-space modeling; and MM4flow~\cite{mm4flow} combines byte streams with transmission patterns. nPrint~\cite{nprint} aligns packet-header bits to a fixed protocol layout, while HF-Transformer~\cite{hf-transformer} serializes selected header fields as a shared byte-unigram sequence. Both regularize header position, but neither assigns typed columns and a separate packet-time axis. ProtoTab instead represents packets as rows, admissible protocol fields as typed columns, and field/time interactions explicitly.

\textbf{Reassessing encrypted-traffic classifiers.}
Wickramasinghe et al.~\cite{sok_etc} show that identity-bearing and session-specific fields can inflate reported performance; Zhao et al.~\cite{pcapencoder} identify complementary inflation from per-packet splitting and encoder-adapting evaluation. In Zhao et al.'s packet-level, per-flow, frozen evaluation, Pcap-Encoder reaches only 13.0 Macro-F1 on TLS-120 without IP addresses. These studies explain several immediate sources of inflated results, but not why performance remains low once those cues are withheld: insufficient retained information or ineffective representation. ProtoTab addresses this residual question with a fixed observer schema and explicit field/time organization.

\section{Method}

\subsection{Framework Overview}

ProtoTab is a protocol-native tabular encoder for the observer record left by encryption and the fixed admissibility filter (P1). It keeps protocol fields as typed table columns and uses parser-defined field identity and protocol type as architectural priors.

\textbf{Flow Definition.} The 5-tuple assembles flows and determines initiator/responder direction. Before model input, IP addresses and ephemeral (dynamic/private) ports are discarded; stable service ports remain, and endpoint identity is reduced to a \texttt{direction} flag.

\textbf{Problem Formulation.} Given a traffic flow $\mathcal{F}$ consisting of $T$ packets $\{p_1, p_2, \ldots, p_T\}$, we extract $N$ protocol fields from each packet, forming a tabular flow representation $\mathbf{X} = [x_i^t]_{T \times N}$ where $x_i^t$ denotes the $i$-th field in packet $t$ as a $T \times N$ multi-row table. Our goal is to learn an encoder $f_\theta: \mathbb{R}^{T \times N} \rightarrow \mathbb{R}^d$ that maps traffic flows to discriminative representations for downstream classification tasks. 

\textbf{Architecture Overview.} As illustrated in Fig.~\ref{fig:workflow}, ProtoTab's architecture consists of four components: (1) typed protocol-field table construction from raw traffic and temporal metadata; (2) the fixed admissibility filter that enforces the retained observer schema for inputs and targets (P1); (3) fixed typed columns and field-position encodings that expose field identity, together with field-specific embeddings that provide field-conditioned value mappings (P2); and (4) a dual-axis Transformer whose \emph{field-axis} attention models cross-field dependencies within a packet (P2), while its \emph{time-axis} attention models cross-packet evolution and consumes the flow-internal inter-packet interval as a first-class field (P3). This interval is available online from consecutive packets of the same flow. For downstream tasks, we evaluate both with frozen encoder (assessing representation quality) and full fine-tuning (assessing end-task performance).

\subsection{Typed Protocol-Field Table Construction}

\textbf{Observation columns.} We construct the input table through a deterministic packet-dissection pipeline, selecting 62 columns from Wireshark's registered frame, IP, TCP, and UDP dissector fields~\cite{wireshark_tshark}. Following prior leakage audits~\cite{sok_etc,pcapencoder}, the observer model excludes payload, identity-bearing channels, raw random/session-state values, and deployment-specific artifacts. The resulting retained schema comprises parser-exposed header fields and capture-time metadata; our implementation retains 23 columns.

\textbf{Flow Sampling.} We set the packet-window hyperparameter to $T=20$ and sample the first 20 packets from each flow. Flows longer than 20 packets are truncated; shorter flows are padded with a mask indicating valid positions. This choice matches the early-stage decision requirement of application classification: many downstream uses (e.g., QoS scheduling and online traffic management) must classify a flow before the full session is observed, while still needing enough packets to cover the TCP/TLS handshake and the first application-dependent burst. The default window is held fixed across datasets rather than tuned per benchmark. Security tasks that target long-horizon attacks may increase $T$ to observe later behavior, but this changes the observation budget and increases both training and inference cost.

\textbf{Typed Field Encoding.} Admissible columns have heterogeneous numeric domains, so each column is normalized according to its protocol type (length, timing, binary flag, or categorical code). The XGBoost table probe consumes these normalized columns directly; nPrint instead uses its native protocol-aligned bit encoding. Reconstruction losses are computed per field and then averaged with equal field weights.

\subsection{Admissibility Filter (P1)}\label{sec:filtering}
 
The schema is determined once by the passive observer model and protocol semantics, before any learning stage. Neither reconstruction errors nor downstream labels are used to select columns. The same retained columns are used unchanged for pretraining, frozen evaluation, fine-tuning, and inference. Byte-level MAE objectives that reconstruct serialized packet inputs can make raw random and non-canonical session-state values part of the target unless those columns are filtered beforehand.
 
\textbf{Admissibility constraints.} We partition parser-exposed columns into admissible ($\mathcal{S}_g$), raw random/session-state ($\mathcal{S}_r$), topology-dependent ($\mathcal{S}_n$), and environment-dependent ($\mathcal{S}_e$) sets:
 
\begin{itemize}
    \item \textit{Raw random/session-state fields} ($\mathcal{S}_r$): fields whose raw values are randomized or session-seeded rather than canonical protocol behavior. This includes \texttt{ip.id}, packet checksums, and absolute TCP sequence/acknowledgment numbers; TCP byte-progress signals are admissible only after per-flow relative normalization.
    
    \item \textit{Topology-dependent fields} ($\mathcal{S}_n$): IP addresses and ephemeral (dynamic/private) ports that encode network topology or endpoint assignment rather than behavior.
    
    \item \textit{Environment-dependent fields} ($\mathcal{S}_e$): fields reflecting OS configurations or transient conditions~\cite{sok_etc}---\texttt{ip.ttl}, \texttt{ip.flags.df}, \texttt{tcp.window\_size}, \texttt{tcp.analysis.ack\_rtt}---that can encode context-specific conditions and thereby support contextual or temporal overfitting.
    
    \item \textit{Admissible fields} ($\mathcal{S}_g$): the retained columns available to the passive encrypted-traffic observer.
\end{itemize}
 
\textbf{Why joint input/target admissibility matters.} Every field admitted to a masked autoencoder also enters its reconstruction objective. Random or session-seeded fields can be difficult to reconstruct from retained context, while raw session state may favor per-flow offsets over relative progress. Fig.~\ref{fig:ablation_selective_masking} is a post hoc diagnostic of inverse-transformed MSE in original units; neither these errors nor their ranking determines schema membership. Adding these fields to inputs and targets lowers frozen Macro-F1 by 38 points on CipherSpectrum (Table~\ref{tab:ablation}); this joint ablation does not isolate the target mechanism.
 
\textbf{Admissibility filter (P1).} Raw fields in $\mathcal{S}_r \cup \mathcal{S}_n \cup \mathcal{S}_e$ are excluded from model inputs and reconstruction targets. When session-state evolution is meaningful, the implementation first applies per-flow de-randomization and treats the canonicalized value as a separate candidate column. Raw fields appear only in the nPrint baseline and the explicit P1 ablation.

\subsection{Field-Specific Embeddings}\label{sec:field_embed}
 
Byte-based methods project all bytes through a shared embedding, so semantically distinct fields receive the same initial value mapping and must be disambiguated by position and context. Byte position provides only an indirect and layout-dependent proxy for field semantics, whereas parser-defined table columns expose field identity explicitly.
 
We assign each retained field column its own embedding function:
\begin{equation}
E_i(x_i^t) = \mathbf{W}_i x_i^t + \mathbf{b}_i, \quad
\mathbf{e}_i^t = E_i(x_i^t) + \mathbf{p}_i + \mathbf{q}_t
\end{equation}
where $\mathbf{W}_i, \mathbf{b}_i$ are specific to retained column $i$, $\mathbf{p}_i$ is field position encoding, and $\mathbf{q}_t$ is temporal position encoding. The fixed column and $\mathbf{p}_i$ expose field identity, $E_i$ supplies its value mapping, and field-axis attention models within-packet interactions (\S\ref{sec:dual_axis_arch}). These are distinct aspects of \textbf{P2}. The P2 ablation replaces all $E_i$ with one shared projector while retaining $\mathbf{p}_i$, isolating the value-mapping aspect while field-position identity remains available.
 
\textbf{Rationale.} Shared projectors pass numerically identical values from semantically different fields through the same value mapping and couple gradients from heterogeneous field types through the same parameters. Field-specific projectors remove this forced parameter sharing and allow each protocol field to learn a separate semantic subspace.

\subsection{Dual-Axis Transformer Architecture}\label{sec:dual_axis_arch}

\textbf{Dual-Axis Attention.} Factorizing attention along orthogonal axes is a well-established technique, originally proposed for high dimensional tensors~\cite{ho2019axial} and widely adopted in video understanding~\cite{bertasius2021space,arnab2021vivit}. Our contribution is its alignment with protocol structure: ProtoTab factorizes attention on $\mathbf{E} \in \mathbb{R}^{T \times N \times d}$ into two axes dictated by the data's tabular structure. \textit{Time-axis attention} models dependencies across $T$ packets for each field position, allowing the model to reason over inter-packet evolution of each typed field---including the flow-internal inter-packet interval, which is available online as packets arrive and is embedded as a first-class temporal field. In byte-form controls, time deltas are ordinary tokens rather than typed temporal fields (P3). TCP header timestamps (TSval/TSecr) cannot substitute, as they reflect sender clocks rather than arrival times per RFC~7323~\cite{rfc7323}. \textit{Field-axis attention} models dependencies across $N$ fields within each packet.

\textbf{Attention-Score Complexity.} Full self-attention over $TN$ packet-field cells costs $O((TN)^2d)$. The time and field axes cost $O(T^2Nd)$ and $O(TN^2d)$, respectively, totaling $O(TN(T{+}N)d)$. For $T{=}20$ and $N{=}23$, this is a theoretical $10.7\times$ reduction in the attention-score term while aligning computation with packet and field structure.
 
\textbf{Encoder Architecture.} The encoder consists of $L$ transformer blocks:
\begin{align}
\mathbf{H}_{\text{time}}^\ell &= \text{TimeAttn}(\text{LN}(\mathbf{H}^{\ell-1})) + \mathbf{H}^{\ell-1} \\
\tilde{\mathbf{H}}^\ell &= \text{FFN}(\text{LN}(\mathbf{H}_{\text{time}}^\ell)) + \mathbf{H}_{\text{time}}^\ell \\
\mathbf{H}_{\text{field}}^\ell &= \text{FieldAttn}(\text{LN}(\tilde{\mathbf{H}}^\ell)) + \tilde{\mathbf{H}}^\ell \\
\mathbf{H}^\ell &= \text{FFN}(\text{LN}(\mathbf{H}_{\text{field}}^\ell)) + \mathbf{H}_{\text{field}}^\ell
\end{align}

\subsection{Training and Downstream Classification}\label{sec:pretrain_downstream}
 
Masked protocol-field prediction initializes the encoder. Let $\mathcal{M}_i$ be the masked, non-padding positions of retained field $i$, and $\mathcal{S}_{+}$ the fields with $|\mathcal{M}_i|>0$ in a mini-batch:

\begin{equation}
\mathcal{L}_{\text{pretrain}}=
\frac{1}{|\mathcal{S}_{+}|}\sum_{i\in\mathcal{S}_{+}}
\frac{1}{|\mathcal{M}_i|}\sum_{(b,t)\in\mathcal{M}_i}
\ell_i\!\left(\hat{\tilde{x}}_{bti},\tilde{x}_{bti}\right).
\end{equation}
Targets $\tilde{x}_{bti}$ are normalized per field and $\ell_i$ is field $i$'s loss. The nested means give equal field weights; the P1 ablation extends $\mathcal{S}_{+}$ with $\mathcal{S}_r$. Section~\ref{sec:component_analysis} separates the margin already present without pretraining from the additional pretraining gain.
 
For downstream tasks, masked mean pooling over time and fields produces $\mathbf{z} \in \mathbb{R}^d$, followed by an MLP head. We evaluate frozen and fully fine-tuned encoders.

\section{Experiments}\label{sec:experiments}

Our experimental evaluation\footnote{\raggedright Anonymous code repository: \url{https://anonymous.4open.science/status/table_transformer-530B}.} is designed to answer three questions:
\begin{enumerate}
    \item[\textbf{Q1.}] How does observer pruning affect modern-encrypted 
    and legacy benchmarks? 
    (\S\ref{sec:sni_analysis})
    \item[\textbf{Q2.}] After matching the observer input and 
    pretraining corpus, does protocol-native modeling outperform byte-form modeling? 
    (\S\ref{sec:main_results})
    \item[\textbf{Q3.}] How do byte encoding, schema enforcement, 
    structured field/time modeling, and masked field pretraining relate to the gains? 
    (\S\ref{sec:component_analysis})
\end{enumerate}

\subsection{Experimental Setup}\label{sec:setup}

\textbf{Pretraining Data.} We use unlabeled MAWI traces from Jan. 1, 2025~\cite{mawi} (137M packets, 9.6GB) for ProtoTab and for re-pretraining the four byte baselines described below. The Japan--US backbone traffic spans thousands of autonomous systems.

\textbf{Downstream Benchmarks.} We evaluate on four datasets with materially different encryption regimes. CipherSpectrum and TLS-120 contain modern encrypted web traffic. Wickramasinghe et al. classify 98.9\% of ISCXVPN2016 sessions and 94.7\% of USTC-TFC2016 sessions as unencrypted, and describe both datasets as dominated by plaintext communications~\cite{sok_etc}. The class and flow counts below describe our evaluated preprocessed data; these percentages describe the source datasets audited in their study.
\begin{itemize}
    \item \textbf{CipherSpectrum\footnote{CipherSpectrum's captures 
    predate widespread ECH deployment and contain plaintext SNI and ALPN; 
    preprocessing removes both to match the retained observer 
    model.}}~\cite{sok_etc}: our evaluated data contains 41 website classes/41K flows, 
    exclusively TLS 1.3 traffic covering the three major TLS 1.3 cipher suites. No 
    unencrypted sessions or deprecated-cipher traffic. \textit{This is our 
    primary benchmark}: after SNI/ALPN removal, it represents an observer 
    view with encrypted payload and no identity-bearing handshake fields.
    \item \textbf{TLS-120 (CSTNET-TLS1.3)}~\cite{et-bert,pcapencoder,sok_etc}: our evaluated data contains 120 website classes/46K flows. Its source dataset contains exclusively encrypted traffic using current TLS~1.3 AEADs; the TLS-120 preprocessing removes the plaintext ClientHello and hence SNI and ALPN~\cite{pcapencoder}. This benchmark tests fine-grained website classification across 120 classes.
    \item \textbf{ISCX-VPN}~\cite{iscx}: our evaluated data contains 16 application classes/16K flows over VPN and non-VPN connections. Wickramasinghe et al.'s audit categorizes 98.9\% of the source dataset's sessions as unencrypted~\cite{sok_etc}.
    \item \textbf{USTC-TFC}~\cite{ustctfc}: our evaluated data contains 20 classes/498K flows, including 10 malware families. The same audit categorizes 94.7\% of the source dataset's sessions as unencrypted~\cite{sok_etc}. We use these two legacy datasets as contrasts; the modern encrypted-observer evidence comes from TLS-120 and CipherSpectrum.
\end{itemize}

\textbf{Retained-Schema Data Preparation.} We follow the leakage-aware preprocessing of Wickramasinghe et al.~\cite{sok_etc} and Zhao et al.~\cite{pcapencoder}: inadmissible observation channels are withheld before learning.
At the \textit{observation-schema level}, we enforce the observer model (\S\ref{sec:observation_model}): (1) withhold IP addresses and ephemeral (dynamic/private) ports but retain stable service ports; (2) mask raw session-specific fields (IP ID, absolute TCP sequence/acknowledgment numbers, checksums) and environment-dependent fields (TTL, window size, TCP timestamps), following prior occlusion methodology~\cite{sok_etc}. Sequence-space information is used only after per-flow relative normalization and the same admissibility filter.
We refer to this combined regime as the \textit{retained-schema} setting throughout.
Payload bytes are excluded from ProtoTab and all matched baseline inputs. In the main results, retained fields are serialized into each baseline's byte, token, or image format without SNI/ALPN, addresses, or ephemeral ports; stable service ports remain.

\textbf{Baselines.} Our pretrained byte baselines are ET-BERT~\cite{et-bert}, YaTC~\cite{yatc}, NetMamba~\cite{netmamba}, and TrafficFormer~\cite{trafficformer}; the header-focused baselines and controls are described below.

\textit{Released-checkpoint byte baselines.} For comparison, Table~\ref{tab:decomposition} reports each prior model initialized from its authors' released checkpoint and fine-tuned using its recommended downstream procedure.

\textit{Same-corpus byte baselines.} We re-pretrain ET-BERT, YaTC, NetMamba, and TrafficFormer from scratch on the same raw unlabeled corpus used by ProtoTab, using each model's native objective. Table~\ref{tab:main_results} freezes these encoders and trains only the downstream classifier, whereas Table~\ref{tab:decomposition} fully fine-tunes them on the same downstream folds and retained inputs.

Two supervised header-byte baselines are trained from scratch: Byte-Packed concatenates native-width header encodings; because no code is released, our HF-Transformer reimplementation~\cite{hf-transformer} follows its published selected-header unigram Transformer under the common payload/identity suppression and 20-packet budget. nPrint~\cite{nprint} uses its aligned header-bit representation with XGBoost; payload, addresses, and ephemeral ports are suppressed, while other native positions remain.

\textbf{Evaluation Protocol.} Following Zhao et al.~\cite{pcapencoder}, folds are constructed at the flow level, with all packets of a bi-flow assigned to the same train/validation/test fold.

We primarily report \textbf{full fine-tuning} and additionally report frozen-encoder transfer. Tables~\ref{tab:main_results}--\ref{tab:decomposition} report mean$\pm$std. over stratified 3-fold cross-validation; the diagnostic and ablation tables report point estimates for their controlled comparisons.

\textbf{Hardware.} Training and evaluation use one NVIDIA A800 80GB GPU and an Intel Xeon Platinum 8374C CPU (18 vCPU @ 2.70\,GHz); Table~\ref{tab:deployment} states the separate preprocessing platform.

\begin{table*}[h]
\centering
\caption{Frozen-encoder transfer from same-corpus checkpoints under the common retained-schema observer model (mean$\pm$std.).}
\label{tab:main_results}
\begin{tabular}{lcccccccc}
\toprule
\multirow{2}{*}{\textbf{Model}} & \multicolumn{2}{c}{\textbf{CipherSpectrum}} & \multicolumn{2}{c}{\textbf{TLS-120}} & \multicolumn{2}{c}{\textbf{ISCX-VPN}} & \multicolumn{2}{c}{\textbf{USTC-TFC}} \\
\cmidrule(lr){2-3} \cmidrule(lr){4-5} \cmidrule(lr){6-7} \cmidrule(lr){8-9}
& Acc & Macro-F1 & Acc & Macro-F1 & Acc & Macro-F1 & Acc & Macro-F1 \\
\midrule
ET-BERT & 2.4$\pm$0.3 & 0.3$\pm$0.2 & 3.4$\pm$0.7 & 1.8$\pm$0.3 & 45.5$\pm$2.1 & 16.3$\pm$1.9 & 70.2$\pm$1.5 & 69.2$\pm$1.8 \\
NetMamba & 7.6$\pm$1.4 & 3.3$\pm$0.9 & 10.0$\pm$1.8 & 5.5$\pm$0.8 & 28.6$\pm$2.0 & 14.0$\pm$1.7 & 31.2$\pm$2.5 & 27.6$\pm$2.8 \\
YaTC & \underline{9.1}$\pm$1.2 & \underline{3.6}$\pm$0.8 & \underline{22.0}$\pm$2.1 & \underline{15.3}$\pm$2.4 & \underline{47.0}$\pm$2.3 & \underline{23.5}$\pm$2.2 & \underline{75.1}$\pm$0.9 & \underline{74.3}$\pm$1.3 \\
TrafficFormer & 5.6$\pm$1.1 & 0.6$\pm$0.3 & 2.4$\pm$0.3 & 1.4$\pm$0.1 & 40.6$\pm$2.5 & 9.1$\pm$2.1 & 59.3$\pm$2.7 & 27.6$\pm$3.2 \\
\midrule
\textbf{ProtoTab} & \textbf{53.4}$\pm$1.3 & \textbf{51.8}$\pm$1.5 & \textbf{55.2}$\pm$1.1 & \textbf{51.3}$\pm$1.4 & \textbf{51.1}$\pm$1.6 & \textbf{42.7}$\pm$1.8 & \textbf{89.1}$\pm$0.6 & \textbf{89.0}$\pm$0.5 \\
\bottomrule
\end{tabular}
\end{table*}

\begin{table*}[h]
\centering
\caption{Fine-tuned performance and controls (mean$\pm$std.). All rows suppress payload and identity inputs; nPrint retains other native header positions. The strongest byte result in each checkpoint group is underlined.}
\label{tab:decomposition}
\begin{tabular}{lcccccccc}
\toprule
\multirow{2}{*}{\textbf{Model}} & \multicolumn{2}{c}{\textbf{CipherSpectrum}} & \multicolumn{2}{c}{\textbf{TLS-120}} & \multicolumn{2}{c}{\textbf{ISCX-VPN}} & \multicolumn{2}{c}{\textbf{USTC-TFC}} \\
\cmidrule(lr){2-3} \cmidrule(lr){4-5} \cmidrule(lr){6-7} \cmidrule(lr){8-9}
& Acc & Macro-F1 & Acc & Macro-F1 & Acc & Macro-F1 & Acc & Macro-F1 \\
\midrule
\multicolumn{9}{l}{\textit{Released-checkpoint byte baselines}} \\
ET-BERT & 11.7$\pm$2.5 & 11.8$\pm$2.3 & 8.8$\pm$1.3 & 3.8$\pm$0.5 & 61.0$\pm$1.9 & 46.7$\pm$2.0 & 92.6$\pm$0.8 & 92.6$\pm$0.5 \\
NetMamba & 34.6$\pm$1.8 & 31.6$\pm$2.5 & \underline{47.8}$\pm$1.6 & \underline{44.7}$\pm$1.3 & 62.4$\pm$1.7 & 51.3$\pm$2.2 & 98.1$\pm$0.1 & 98.0$\pm$0.3 \\
YaTC & \underline{42.3}$\pm$1.7 & \underline{40.0}$\pm$2.1 & 46.0$\pm$2.4 & 42.4$\pm$1.5 & 64.6$\pm$2.2 & \underline{52.3}$\pm$1.9 & \underline{98.3}$\pm$0.3 & \underline{98.2}$\pm$0.4 \\
TrafficFormer & 3.2$\pm$0.8 & 2.0$\pm$2.4 & 23.3$\pm$2.6 & 15.8$\pm$1.7 & \underline{64.7}$\pm$2.4 & 51.5$\pm$2.1 & 96.8$\pm$0.4 & 86.4$\pm$0.8 \\
\midrule
\multicolumn{9}{l}{\textit{Same-corpus byte baselines}} \\
ET-BERT & 12.07$\pm$4.5 & 10.10$\pm$4.7 & 29.67$\pm$3.2 & 27.08$\pm$3.5 & 52.68$\pm$2.3 & 32.01$\pm$2.7 & 78.53$\pm$1.3 & 66.09$\pm$1.6 \\
NetMamba & 57.02$\pm$2.7 & 57.56$\pm$2.8 & 52.18$\pm$2.5 & 50.02$\pm$2.6 & 43.61$\pm$2.8 & 33.12$\pm$3.1 & 92.06$\pm$0.6 & 86.26$\pm$0.9 \\
YaTC & \underline{61.73}$\pm$2.3 & \underline{62.56}$\pm$2.4 & \underline{66.93}$\pm$1.8 & \underline{64.77}$\pm$1.9 & \underline{63.88}$\pm$1.8 & \underline{59.75}$\pm$2.1 & 94.91$\pm$0.4 & 92.93$\pm$0.5 \\
TrafficFormer & 18.27$\pm$3.8 & 14.33$\pm$3.9 & 48.93$\pm$2.6 & 45.40$\pm$2.7 & 59.71$\pm$2.1 & 48.33$\pm$2.5 & \underline{96.14}$\pm$0.3 & \underline{96.91}$\pm$0.3 \\
\midrule
\multicolumn{9}{l}{\textit{From-scratch header-byte baselines}} \\
Byte-Packed Fields & 10.41$\pm$0.75 & 6.45$\pm$0.53 & 2.7$\pm$0.9 & 0.6$\pm$0.3 & 24.8$\pm$2.3 & 24.9$\pm$2.5 & 60.9$\pm$2.1 & 55.7$\pm$2.4 \\
HF-Transformer & 34.9$\pm$2.0 & 32.3$\pm$2.2 & 38.0$\pm$1.9 & 32.3$\pm$2.1 & 27.4$\pm$2.1 & 28.5$\pm$2.3 & 60.8$\pm$1.8 & 54.4$\pm$2.0 \\
\midrule
\multicolumn{9}{l}{\textit{Protocol-structured controls and ProtoTab}} \\
XGBoost Table Probe & 71.6$\pm$0.9 & 71.2$\pm$1.0 & 67.1$\pm$1.1 & 65.0$\pm$1.2 & 54.5$\pm$1.3 & 53.8$\pm$1.4 & 84.8$\pm$0.8 & 90.7$\pm$0.7 \\
nPrint & 67.8$\pm$1.2 & 67.8$\pm$1.3 & 65.4$\pm$1.4 & 60.9$\pm$1.5 & 53.3$\pm$1.5 & 50.6$\pm$1.6 & 81.9$\pm$1.1 & 85.5$\pm$1.0 \\
FT-Transformer (flat) & 77.6$\pm$1.3 & 78.1$\pm$1.2 & 78.1$\pm$1.2 & 76.2$\pm$1.4 & 37.8$\pm$2.3 & 42.6$\pm$2.5 & 79.1$\pm$1.5 & 84.1$\pm$1.3 \\
No pretrain (field arch) & 82.2$\pm$1.0 & 81.7$\pm$1.4 & 82.3$\pm$2.6 & 79.1$\pm$1.3 & 78.6$\pm$2.0 & 54.6$\pm$1.8 & 96.9$\pm$0.5 & 95.9$\pm$0.5 \\
\textbf{ProtoTab} & \textbf{89.6}$\pm$0.8 & \textbf{89.3}$\pm$1.3& \textbf{86.2}$\pm$1.4 & \textbf{83.8}$\pm$0.9 & \textbf{83.6}$\pm$1.8 & \textbf{68.5}$\pm$1.5 & \textbf{98.4}$\pm$0.5 & \textbf{98.4}$\pm$0.3 \\
\bottomrule
\end{tabular}
\end{table*}

\subsection{Observer-Regime Check (Q1)}\label{sec:sni_analysis}

\begin{table}[h]
\centering
\caption{Input-pruning dependency under controlled fine-tuning (Macro-F1 point estimates). $\Delta$: Raw-to-retained change.}
\label{tab:sni_analysis}
\begin{tabular}{l ccc ccc}
\toprule
& \multicolumn{3}{c}{\textbf{USTC-TFC}} & \multicolumn{3}{c}{\textbf{ISCX-VPN}} \\
\cmidrule(lr){2-4} \cmidrule(lr){5-7}
\textbf{Model} & \textbf{Raw} & \textbf{Ret.} & \textbf{$\Delta$} & \textbf{Raw} & \textbf{Ret.} & \textbf{$\Delta$} \\
\midrule
ET-BERT       & 95.7 & 92.6 & $-3.1$ & 54.3 & 46.7 & $-7.6$ \\
NetMamba      & 98.2 & 98.0 & $-0.2$ & 48.6 & 51.3 & $+2.7$ \\
YaTC          & 98.1 & 98.2 & $+0.1$ & 54.8 & 52.3 & $-2.5$ \\
TrafficFormer & 92.9 & 86.4 & $-6.5$ & 49.2 & 51.5 & $+2.3$ \\
\midrule
ProtoTab$^\dagger$ & -- & $\mathbf{98.4}$ & -- & -- & $\mathbf{68.5}$ & -- \\
\midrule
& \multicolumn{3}{c}{\textbf{TLS-120}} & \multicolumn{3}{c}{\textbf{CipherSpectrum}} \\
\cmidrule(lr){2-4} \cmidrule(lr){5-7}
\textbf{Model} & \textbf{Raw} & \textbf{Ret.} & \textbf{$\Delta$} & \textbf{Raw} & \textbf{Ret.} & \textbf{$\Delta$} \\
\midrule
ET-BERT       & 51.5 &  3.8 & $-47.7$ & 26.0 & 11.8 & $-14.2$ \\
NetMamba      & 76.0 & 44.7 & $-31.3$ & 80.1 & 31.6 & $-48.5$ \\
YaTC          & 74.8 & 42.4 & $-32.4$ & 69.3 & 40.0 & $-29.3$ \\
TrafficFormer & 69.2 & 15.8 & $-53.4$ & 32.6 &  2.0 & $-30.6$ \\
\midrule
ProtoTab$^\dagger$ & -- & $\mathbf{83.8}$ & -- & -- & $\mathbf{89.3}$ & -- \\
\bottomrule
\end{tabular}
\vspace{1pt}
\parbox{\columnwidth}{\footnotesize\raggedright $^\dagger$ Raw fields are inadmissible ProtoTab inputs under P1; Table~\ref{tab:ablation} reports the related P1 stress test.\par}
\end{table}

Table~\ref{tab:sni_analysis} validates the benchmark regimes under a common intervention. The retained condition applies the full input pruning defined in \S\ref{sec:setup}, including payload, identity-bearing channels, raw session state, and environment-specific fields. Raw and retained variants use identical initialization and are fine-tuned separately, following prior occlusion methodology~\cite{sok_etc,pcapencoder}.

The two legacy datasets are comparatively insensitive to this intervention: the Raw-to-retained change spans $-6.5$ to $+0.1$ Macro-F1 points on USTC-TFC and $-7.6$ to $+2.7$ on ISCX-VPN. Every byte baseline drops on the modern encrypted datasets, by 31.3--53.4 points on TLS-120 and 14.2--48.5 on CipherSpectrum. This division agrees with the independent audit of their source datasets: USTC-TFC and ISCX-VPN are overwhelmingly unencrypted at the session level, whereas the sources of TLS-120 and CipherSpectrum contain modern encrypted traffic~\cite{sok_etc}.

Payload bytes are absent from every retained result. The stability of the legacy benchmarks therefore shows that protocol and flow-shape cues remain discriminative after full retained-schema pruning in those captures. The large drops across the evaluated models on TLS-120 and CipherSpectrum characterize the harder retained-observer setting studied here, in which classification relies on residual header, direction, and capture-time observations. Q2 asks how best to represent exactly that record.

\begin{figure}[h]
    \centering
    \includegraphics[width=0.8\linewidth]{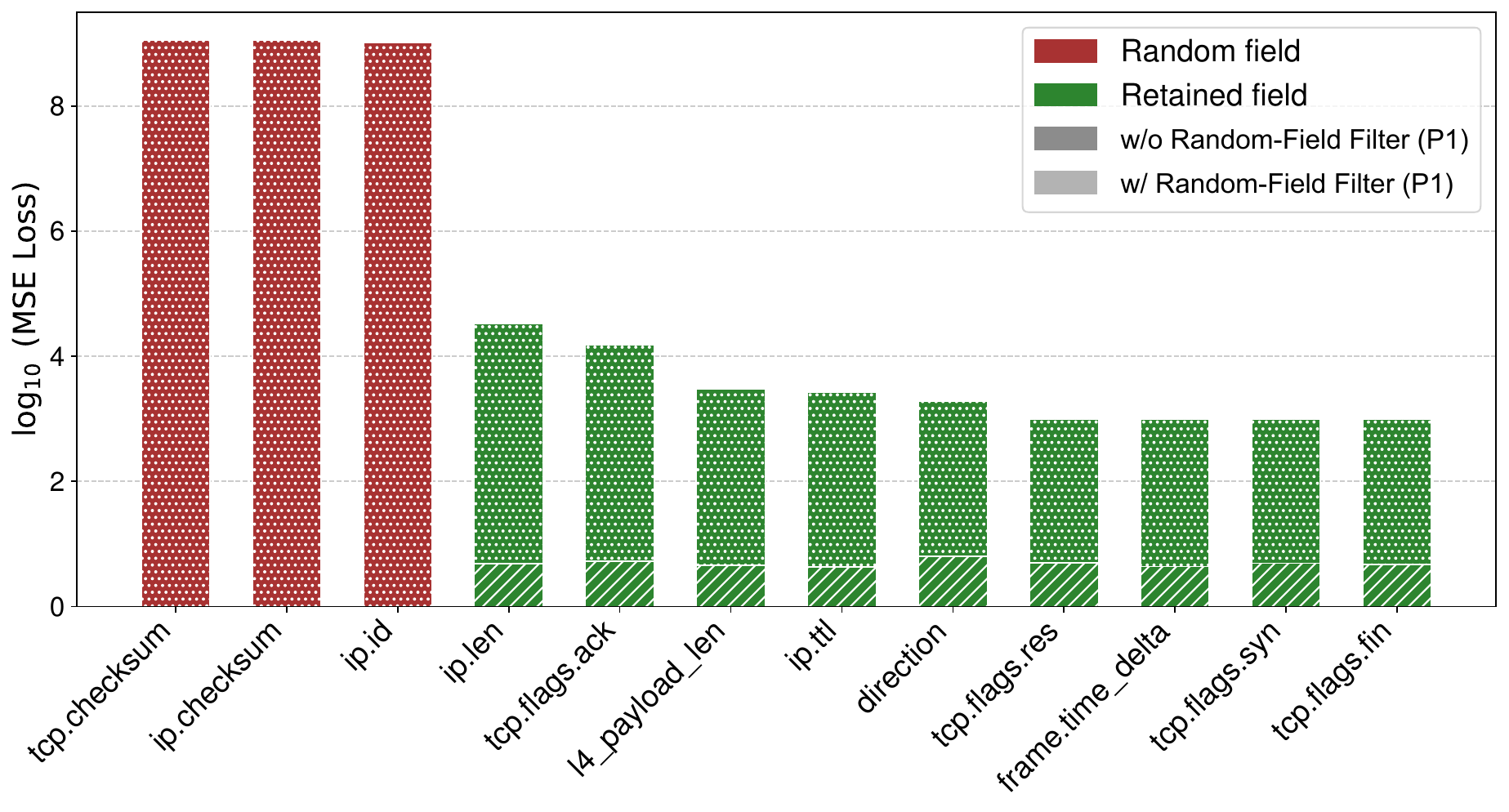}
    \caption{Diagnostic only: inverse-transformed per-field reconstruction MSE for the P1 ablation (log scale, original field units). These errors do not select the schema; training uses independently normalized, equally weighted per-field losses.}
    \label{fig:ablation_selective_masking}
\end{figure}

\begin{figure*}[!t]
    \centering
    \includegraphics[width=0.9\textwidth]{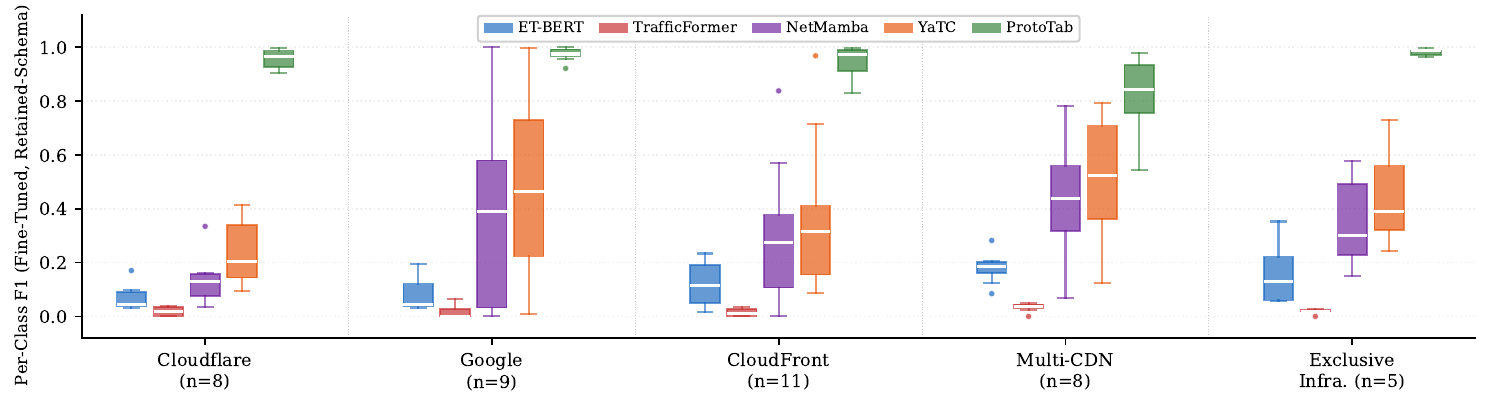}
    \caption{Per-class F1 distributions by CDN provider on CipherSpectrum after end-to-end fine-tuning.}
    \label{fig:boxplot}
\end{figure*}

\begin{figure}[h]
    \centering
    \includegraphics[width=0.82\linewidth]{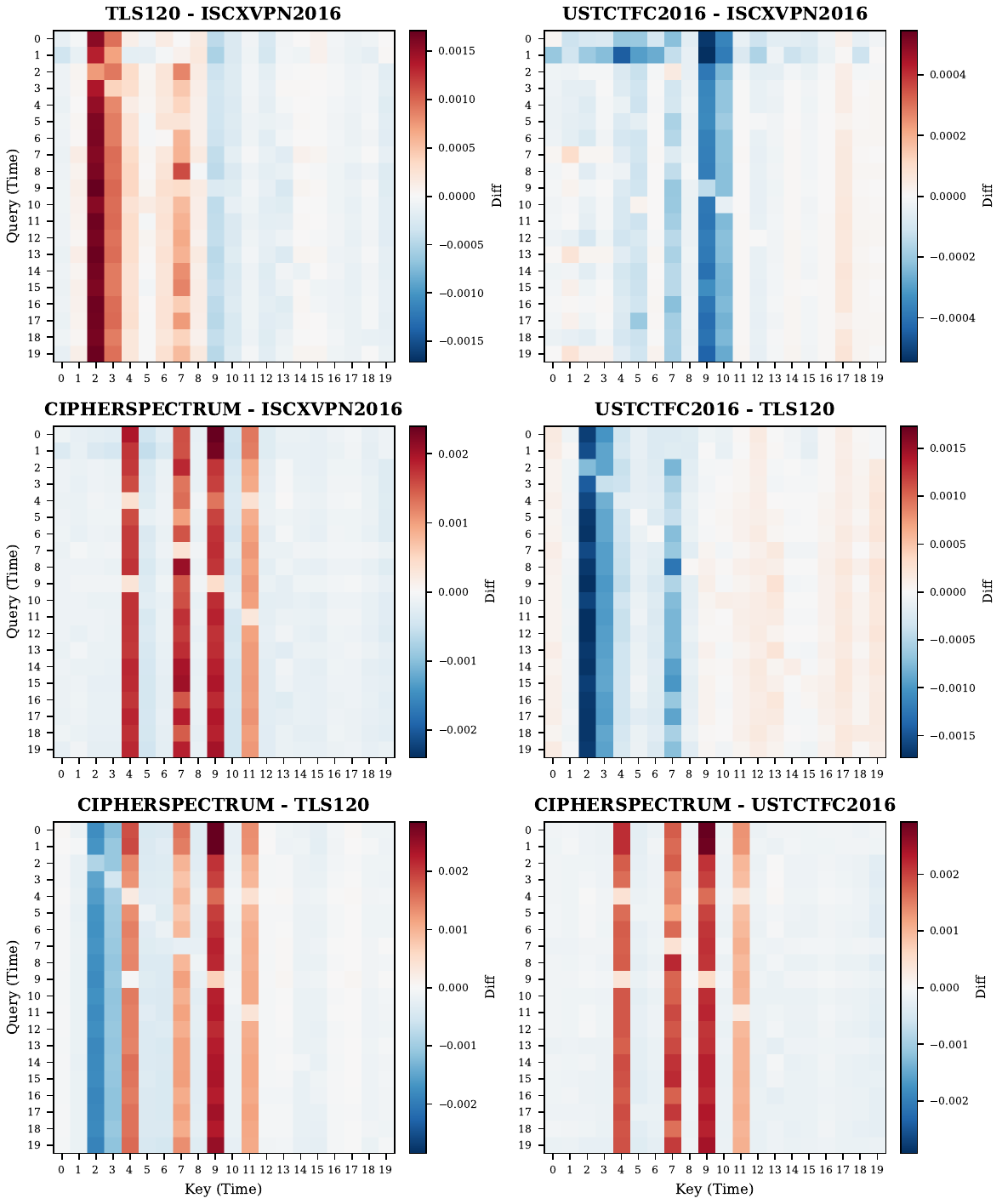}
    \caption{Pairwise time-axis attention differences (first-named minus second-named dataset).}
    \label{fig:time_attention_diff}
\end{figure}

\begin{figure}[h]
    \centering
    \includegraphics[width=0.82\linewidth]{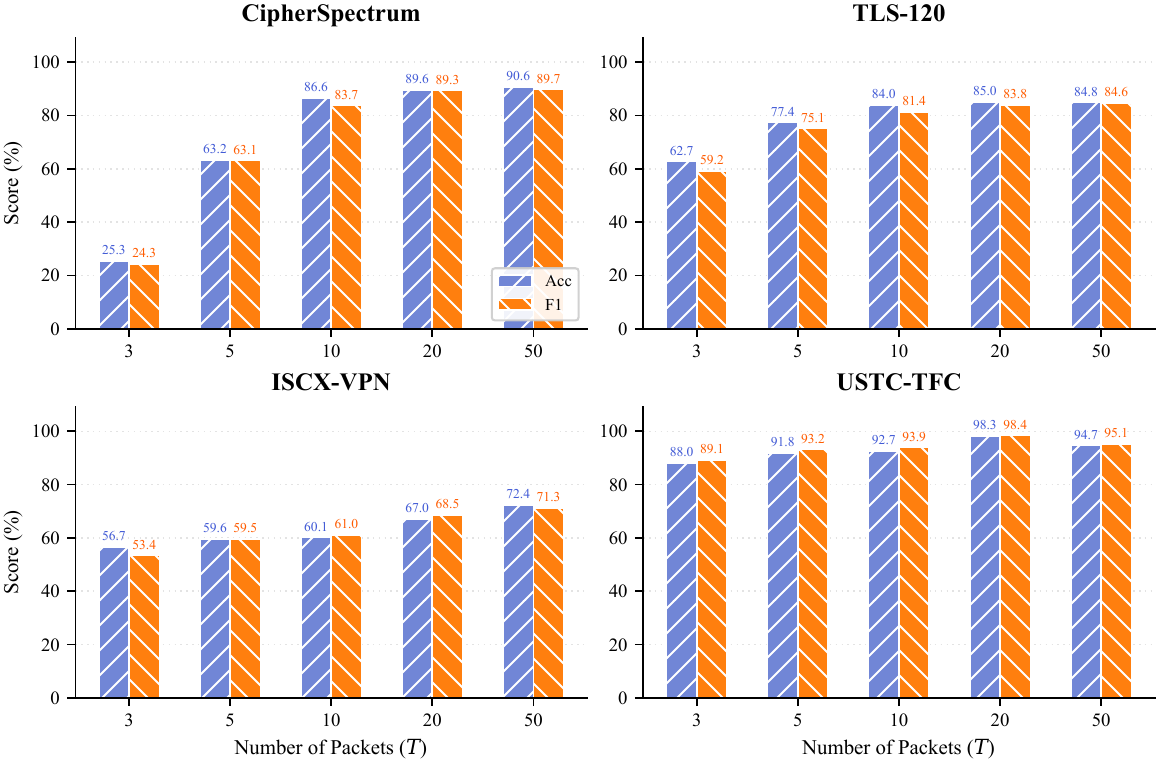}
    \caption{Accuracy and Macro-F1 vs.\ packet count, showing dataset-dependent sensitivity to the observation window.}
    \label{fig:sensitivity_T}
\end{figure}

\subsection{Main Results (Q2)}\label{sec:main_results}

Q2 compares models under the retained schema: every model receives the same protocol fields and capture-time metadata, with payload and identity-bearing fields excluded.

\textbf{Re-pretrained frozen byte encoders remain weak on the two modern encrypted benchmarks.} With all encoders frozen after pretraining on the same raw unlabeled corpus (Table~\ref{tab:main_results}), the strongest evaluated byte model reaches 3.6\% Macro-F1 on CipherSpectrum and 15.3\% on TLS-120, while ProtoTab reaches 51.8\% and 51.3\%. The contrast is smaller on USTC-TFC, whose source dataset is dominated by sessions categorized as unencrypted in Wickramasinghe et al.'s audit. Within this matched setting, frozen byte-form features are least effective on the two modern encrypted benchmarks.

\textbf{ProtoTab's advantage is largest on modern encrypted benchmarks.} Under matched observer inputs and the same raw unlabeled corpus (Table~\ref{tab:decomposition}), ProtoTab exceeds the strongest byte baseline by 26.74 Macro-F1 points on CipherSpectrum and 19.03 on TLS-120, compared with 8.75 on ISCX-VPN and 1.49 on USTC-TFC. The larger modern-encrypted margins provide the central cross-dataset support for protocol-native modeling; Q3 examines how the component controls relate to them.

\textbf{Byte-form performance remains architecture dependent.} Under the matched setting, byte-model Macro-F1 spans 10.10--62.56\% on CipherSpectrum and 27.08--64.77\% on TLS-120. With observer inputs, the raw unlabeled corpus, and folds controlled, this spread quantifies the variation across architectures and native objectives on the same retained record.

YaTC is strongest among the byte models on both modern benchmarks, while NetMamba is competitive on CipherSpectrum but falls further behind on TLS-120. ProtoTab is the only learned model above 80\% Macro-F1 on both. Because the two benchmarks differ in class count, infrastructure, and packet-position emphasis (Fig.~\ref{fig:time_attention_diff}), the result is not confined to a single evaluated dataset.

The narrower legacy-dataset margins do not contradict this result. Q1 shows that their retained header and flow-shape cues are already readily discriminative, leaving less headroom for a representation change.

\subsection{Component Analysis (Q3)}\label{sec:component_analysis}

Table~\ref{tab:decomposition} separates matched pretrained byte baselines, prior header-focused representations (HF-Transformer and nPrint), and same-schema architecture/pretraining controls.

\textbf{HF-Transformer substantially improves over naive header-byte packing on modern encrypted traffic.} Both methods are supervised from scratch without payload or pretraining. HF-Transformer exceeds Byte-Packed Fields by 25.9 Macro-F1 points on CipherSpectrum and 31.7 on TLS-120; the differences are $+3.6$ on ISCX-VPN and $-1.3$ on USTC-TFC. Because field selection and sequence construction both change, this is a method comparison rather than a component ablation.

Yet HF-Transformer remains at 32.3\% on both modern benchmarks, versus ProtoTab's 89.3\% and 83.8\%. Its selected-header byte sequence does not close the observed gap; the same-schema controls next examine typed mappings and structured field/time modeling.

The XGBoost table probe supplies a complementary check that the retained table contains accessible class signal. It reaches 71.2\% on CipherSpectrum, 8.64 points above the best matched byte model; on TLS-120 it is effectively tied with YaTC (65.0\% versus 64.77\%), and on the two legacy datasets it trails the strongest matched byte model. Its evidence is therefore strongest on CipherSpectrum, while the same-schema neural controls provide a more direct within-schema architectural comparison.

\textbf{The field architecture improves over the flat control on all four datasets.} Moving from flat FT-Transformer to the no-pretraining field architecture raises Macro-F1 by 3.6, 2.9, 12.0, and 11.8 points on CipherSpectrum, TLS-120, ISCX-VPN, and USTC-TFC. Both rows use the same field table; the change is associated with the complete structured architecture.

\textbf{The no-pretraining field architecture retains most of the modern-benchmark margin.} Relative to the strongest matched byte baseline, ProtoTab without pretraining is higher by 19.14 and 14.33 Macro-F1 points on CipherSpectrum and TLS-120, respectively, compared with the full model's 26.74- and 19.03-point margins. Masked field pretraining adds 7.60 and 4.70 points within ProtoTab.

\textbf{nPrint provides a direct protocol-aligned baseline.} With XGBoost, it reaches 67.8, 60.9, 50.6, and 85.5\% Macro-F1, respectively---3.4, 4.1, 3.2, and 5.2 points below the typed XGBoost table probe. Because encoding and native header positions both change, this is not a P1 ablation. Fixed protocol alignment alone nevertheless remains below the typed field/time models in this evaluation.

\subsubsection{Ablations Tied to P1, P2, and P3}\label{sec:ablation}

\begin{table}[h]
\centering
\caption{Frozen-encoder ablation of ProtoTab components (point estimates).}
\label{tab:ablation}
\begin{tabular}{lcccc}
\toprule
\multirow{2}{*}{\textbf{Variant}} & \multicolumn{2}{c}{\textbf{CipherSpectrum}} & \multicolumn{2}{c}{\textbf{TLS-120}} \\
\cmidrule(lr){2-3} \cmidrule(lr){4-5}
& Acc & Macro-F1 & Acc & Macro-F1 \\
\midrule
\textbf{ProtoTab (full)} & \textbf{53.4} & \textbf{51.8} & \textbf{55.2} & \textbf{51.3} \\
\midrule
\textit{w/o} Random-Field Filter (P1) & 16.9 & 13.8 & 27.9 & 17.3 \\
\textit{w/o} Field-Spec Embed (P2) & 47.0 & 44.0 & 40.8 & 36.5 \\
\textit{w/o} Temporal Data (P3) & 26.4 & 25.6 & 45.3 & 30.5 \\
\bottomrule
\end{tabular}
\end{table}

\textbf{Random-field filter (P1 test).} Removing the random-field component of the fixed admissibility filter adds random/session-state fields to both the encoder input and masked targets, reducing frozen Macro-F1 by 38.0 points on CipherSpectrum and 34.0 on TLS-120. The schema was fixed before learning; Fig.~\ref{fig:ablation_selective_masking} reports only post hoc, inverse-transformed diagnostic errors. Training uses independent per-field normalization and equal loss weights. These results show that jointly admitting the fields as inputs and targets is harmful in the evaluated ProtoTab configuration; they do not separate the input and target effects.

\textbf{Field-specific value mappings (P2).} Replacing the field-specific projectors with a shared mapping, while retaining fixed field-position identity, drops Macro-F1 by 7.8 points on CipherSpectrum and 14.8 on TLS-120. Within this ProtoTab ablation, field position already identifies the column; the result is consistent with a benefit from field-specific value mappings even when position is known. The larger TLS-120 drop is consistent with its 120-way fine-grained task.

\textbf{Temporal metadata (P3).} Removing flow-internal inter-packet intervals from ProtoTab drops Macro-F1 by 26.2 points on CipherSpectrum and 20.8 on TLS-120. This ablation measures the value of capture-time intervals within ProtoTab on the two modern encrypted datasets. Dual-axis factorization is motivated separately by its protocol-aligned attention-score complexity (\S\ref{sec:dual_axis_arch}); the attention analysis below is descriptive.

Across both modern datasets, removing the random-field filter (P1 test) causes the largest frozen drop, followed by temporal metadata (P3) and field-specific mappings (P2). The relative magnitudes differ: P2 drops 14.8 points on TLS-120 versus 7.8 on CipherSpectrum, whereas removing timing costs 20.8 and 26.2 points. Together with the later attention and window diagnostics, this pattern suggests that field-conditioned mappings matter relatively more on TLS-120, while CipherSpectrum is more sensitive to flow-internal timing.

\subsection{Breadth and Temporal Structure of the Gain}\label{sec:additional_observations}

We next test whether the CipherSpectrum gain persists across shared-infrastructure groups and whether the temporal evidence follows the distinct traffic regimes established in Q1.

\textbf{Shared-infrastructure separation (Fig.~\ref{fig:boxplot}).} CipherSpectrum classes are also grouped by CDN provider. ProtoTab maintains high per-class F1 across Cloudflare, Google, CloudFront, multi-CDN, and exclusive-infrastructure groups.

The Cloudflare group is a stress case because its classes share reverse-proxy infrastructure and similar handshake parameters. High F1 within that group shows that the aggregate result is not concentrated in classes with exclusive infrastructure; this remains an in-dataset breadth check rather than a provider-held-out generalization test.

\textbf{Dataset-dependent temporal evidence (Fig.~\ref{fig:time_attention_diff}).} The pairwise maps show TLS-120 placing relatively more attention on early-packet positions than the compared datasets, consistent with its diverse handshakes. CipherSpectrum places relatively more on later positions, where post-handshake size and timing remain available among classes sharing CDN infrastructure. Contrasts involving USTC-TFC and ISCX-VPN are more diffuse. These are relative cross-dataset differences, not absolute attention maps; the P3 ablation in Table~\ref{tab:ablation} supplies the controlled evidence that flow-internal timing matters.

\textbf{Packet-window sensitivity (Fig.~\ref{fig:sensitivity_T}).} CipherSpectrum benefits most from observing beyond the initial packets, rising from 24.3\% Macro-F1 at $T{=}3$ to 83.7\% at $T{=}10$ and saturating near $T{=}20$. TLS-120 starts higher, matching its relative early-packet emphasis in Fig.~\ref{fig:time_attention_diff}. ISCX-VPN's improvement over longer windows is consistent with VPN and non-VPN flow shapes unfolding, while USTC-TFC peaks near $T{=}20$, where most evaluated flows are already covered.

The default $T{=}20$ is therefore an early-stage packet budget for the evaluated application-classification tasks. In these data, it covers the handshake and the first application-dependent burst for encrypted services. Tasks that depend on delayed behavior, such as long-dwell attacks or late command-and-control exchanges, may choose a larger $T$; this changes the observation budget and increases the encoder cost $O(TN(T{+}N)d)$ without changing the protocol-field schema.

The flow-length distributions help explain the different curves. In our evaluated USTC-TFC data, 87.9\% of flows contain at most ten packets and 99.4\% contain at most twenty; larger windows mainly add padding positions and coincide with lower performance at $T{=}50$. Nearly all evaluated CipherSpectrum flows exceed ten packets. A $T{=}3$ window captures little beyond the initial exchange, while $T{=}10$ includes the first application-dependent records and coincides with the sharp jump in Fig.~\ref{fig:sensitivity_T}. Thus $T{=}20$ is a robust observation budget for the evaluated early-stage tasks.

\subsection{Deployment Cost}\label{sec:deployment}

\begin{table}[h]
\centering
\caption{Model size and raw throughput. Neural inference uses the A800; XGBoost uses the CPU; ``n/a'' denotes unprofiled stages.}
\label{tab:deployment}
\begin{tabular}{lrrr}
\toprule
\textbf{Model} & \textbf{Params} & \textbf{Infer.} & \textbf{Preproc.} \\
 & (M) & (flow/s) & (flow/s) \\
\midrule
ET-BERT & 108.86 & 300.7 & n/a \\
TrafficFormer & 108.87 & 132.1 & n/a \\
YaTC & 2.33 & 1767.3 & 148.1 \\
NetMamba & 2.17 & 1759.5 & 148.1 \\
XGBoost Table Probe & 0.02 & 14791.1 & n/a \\
\midrule
ProtoTab & 13.14 & 380.8 & 1379.6 \\
\bottomrule
\end{tabular}
\end{table}

Table~\ref{tab:deployment} separates inference from preprocessing; parameter counts follow each implementation. ProtoTab uses 13.14M parameters, runs at 380.8 flow/s, and its TShark field extraction reaches 1379.6 flow/s on an Intel(R) Xeon(R) w5-3423 CPU at 2.1\,GHz.

The TShark rate is not a preprocessing ceiling: optimized libpcap-based or hardware-accelerated packet processing was not evaluated. YaTC and NetMamba run at about 1760 flow/s, but their byte-to-image conversion reaches 148.1 flow/s.

XGBoost Table Probe is the fastest measured classifier at 14{,}791.1 flow/s on CPU, but its Macro-F1 is 71.2\% on CipherSpectrum and 65.0\% on TLS-120, versus ProtoTab's 89.3\% and 83.8\%. The result exposes an accuracy--compute trade-off.

\section{Conclusion and Limitation}\label{sec:conclusion}

\textbf{Conclusion.} Under the fixed retained observer schema, results across four benchmarks show that representation affects how available evidence is learned. ProtoTab organizes retained protocol and timing information as a typed packet-by-field table with a joint input/target schema, per-column mappings, and dual-axis attention. With inputs and the raw unlabeled corpus matched, it exceeds the strongest matched byte baselines by 26.74 and 19.03 Macro-F1 points on CipherSpectrum and TLS-120. Matching inputs rules out additional observation channels; the controls associate the margins with its architecture, schema, mappings, timing, and pretraining.

\textbf{Limitation.} Findings are limited to this schema, the evaluated captures, and early-flow windows. Protocol or parser changes require a renewed schema audit. Our folds and CDN analysis do not test cross-network, temporal, regional, or provider transfer, and $T{=}20$ omits late-flow evidence.

\clearpage
\bibliographystyle{IEEEtran}
\bibliography{sample-base}

\end{document}